\documentclass[aps,twocolumn,floatfix]{revtex4-1}
\usepackage{epsfig}
\usepackage{amssymb}
\usepackage{amsmath}
\usepackage{amsfonts}
\usepackage{mathtools}
\usepackage[overload]{empheq}
\usepackage{graphicx}
\usepackage{color}
\usepackage{microtype}

\begin{document}

\title{Reversible compression of an optical piston through Kramers dynamics}
\author{Gabriel Schnoering}
\author{Cyriaque Genet}
\email[Electonic address: ]{genet@unistra.fr}
\affiliation{ISIS \& icFRC, Universit\'{e} de Strasbourg and CNRS
(UMR 7006), 8 all\'{e}e Gaspard Monge, F-67000 Strasbourg, France.}

\begin{abstract}
We study the reversible crossover between stable and bistable phases of an over-damped Brownian bead inside an optical piston. The interaction potentials are solved developing a method based on Kramers' theory that exploits the statistical properties of the stochastic motion of the bead. We evaluate precisely the energy balance of the crossover. We show that the deformation of the optical potentials induced by the compression of the piston is related to a production of heat which measures the non-adiabatic character of the crossover. This reveals how specific thermodynamic processes can be designed and controlled with a high level of precision by tailoring the optical landscapes of the piston.
\end{abstract}

\maketitle

{
Optically trapped Brownian particles constitute ideal test systems for non-equilibrium statistical physics with a great variety of stochastic protocols under external force fields that can be implemented \cite{SeifertRPP2012}. A particular attention has been devoted to measuring thermal fluctuation-induced escape over an optical potential barrier and exploring Kramers rate theory, including the observation of stochastic synchronization \cite{SimonPRL1992,McCannNat1999,HayashiOptComm2008,SilerNJP2010}. More recently, quantitative tests of so-called fluctuation theorems have involved optically trapped nanoparticles, both in the over- and under-damped regimes \cite{WangPRL2002,BustamantePhysToday2005,BlicklePRL2006,GieselerNatNano2014}. Bistable optical potentials are currently exploited for developing Szilard-types engines and studying the connections between information theory and thermodynamics \cite{BerutNat2012,RoldanNatPhys2014,JunPRL2014}. 

In this Letter, we monitor, at room temperature, the crossover between stable and bistable motions of a thermalized over-damped Brownian particle optically trapped in front of a mirror. For specific positions of the mirror, the coherent superposition of the incident trapping beam and the reflected beam induces dynamical bistability where the particle is activated between two distinct positions along the optical axis. We demonstrate that the whole interaction potential can be solved by interpreting the two positions as distinguishable metastable states. Diffusion limited escape rates and associated activation energies are extracted, together with the actual distance separating the metastable states. Remarkably, this is performed without any preliminary spatial calibration of our optical setup.

While the instantaneous position of the particle is a stochastic process, the position of the mirror is an external variable that controls the optical force field applied to the particule. We show that the movable mirror acts as an optical piston that quasi-statically injects reversible work into the system in the form of Helmholtz free energy. From this description, the energy cost on the Brownian particle associated with the compression of the piston can be measured precisely. This reveals that the crossover is close to the adiabatic limit. We show that the deformation of the optical potentials through the displacement of the piston produces some reversible heat that fully accounts for this small deviation from perfect adiabaticity. Remarkably, this relation between the optical landscapes and the production of heat points to an efficient resource for designing specific thermodynamic processes.

\begin{figure}[h!]
\includegraphics[width=0.50\textwidth]{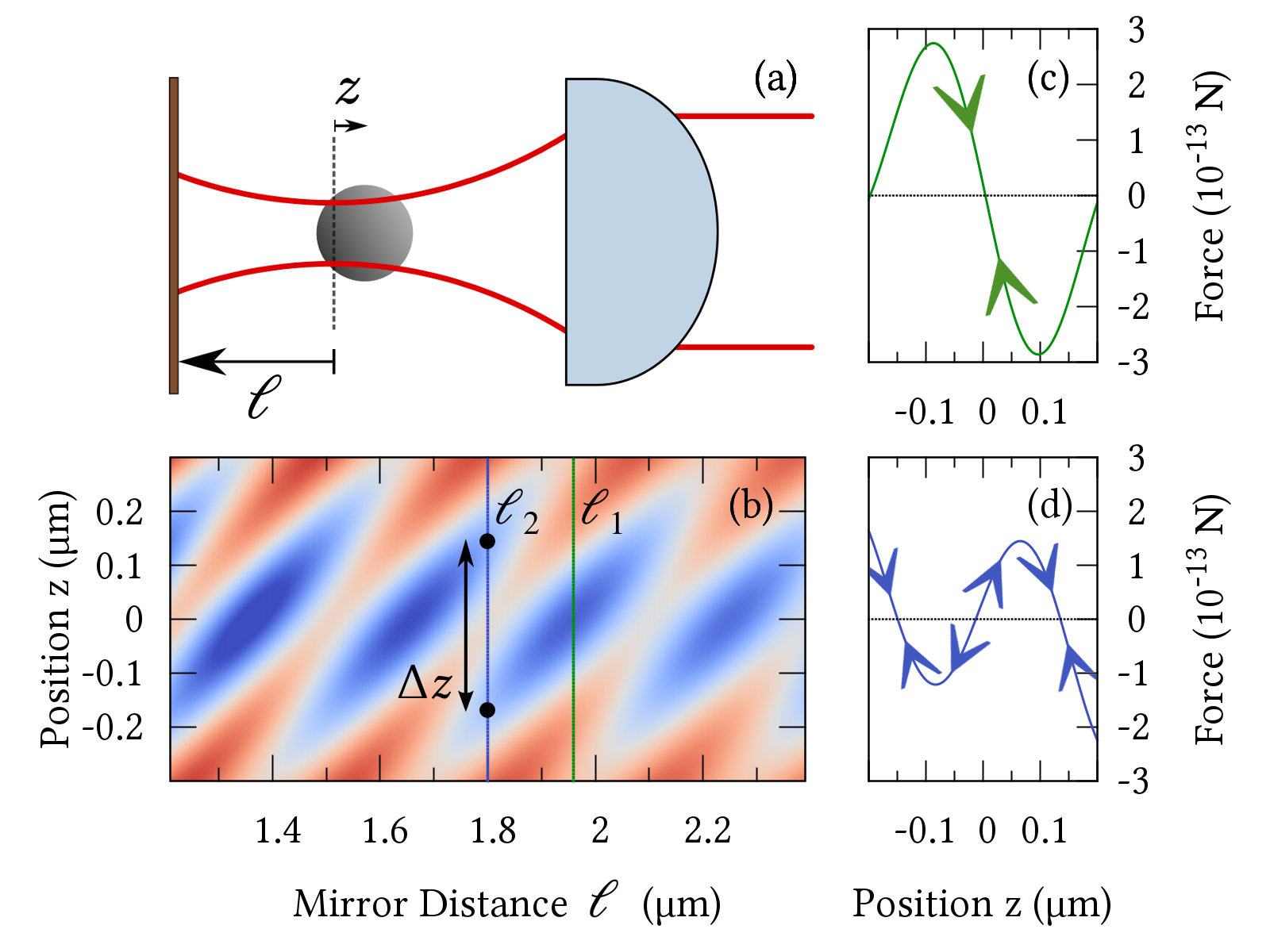}
 \caption{(a) Sketch of the trapping configuration. A micron-size bead (radius $R=500$ nm) is trapped by a Gaussian beam ($\lambda=785$ nm) focused through a microscope objective. The position of the bead $z$ is defined with respect to the beam waist $w_0$. The end-mirror is positioned at a distance $\ell$ from the waist. (b) Evolution of the optical landscape in the vicinity of the waist ($z=0$) and as a function of $\ell$. Blue colors correspond to regions of higher intensity, i.e. deeper potential energy and thus to stable positions. The case of a stable landscape corresponds to the distance $\ell_1$ (green line) and a bistable landscape is crossed at $\ell_2$ (blue line). Plots of the force diagrams associated with $\ell_1$-stability (c) and $\ell_2$-bistability (d). 
}
 \label{figure_scheme}
\end{figure}

In our experiment, a single polystyrene bead is optically trapped by a focused Gaussian beam in a water cell at a typical $2~\mu$m distance from a metallic mirror (see details in Appendix \ref{AppA}). The trapping beam, characterized by a fixed waist $w_0$ located at $z=0$, propagates in the fluid along the $z>0$ optical axis with a wave vector $+k$ -see Fig. 1 (a). It is $M\times$ magnified through the transparent bead acting as a lens and reflected with a reflection amplitude $r[\lambda]$ by the mirror placed at a distance $\ell$ from $w_0$.  This creates a coherent optical landscape at the position of the bead $z$ measured from the waist (see Appendix \ref{AppB1})
\begin{eqnarray}
{I}^{\rm opt}_{M}(z,\ell)=|{\bf E}^{+k}(z)+r[\lambda]\cdot {\bf E}^{-k}_{M}(z-2\ell)|^{2},
\end{eqnarray}
displayed in Fig. 1 (b) as a function of $z$ and $\ell$ for a fixed value of $M$. As expected from its interfering nature, the optical landscape profile changes with the waist-mirror distance $\ell$.

The corresponding evolution has direct consequences on the dynamics of the trapped bead in the vicinity of the waist. Although the approach we develop below is fully general, these consequences are most easily described in a dipolar approach. Here, the non-absorbing bead is modeled by a real dipolar polarizability $\alpha$ and the optical interaction potential is $U_M (z,\ell)=-A\cdot\alpha\cdot {I}^{\rm opt}_{M}(z,\ell) /(2\varepsilon_0 n_2 c)$, with $n_2$ the refractive index of the fluid (see Appendix \ref{AppB2}). The coupling constant $A$ allows accounting for bead size effects, with $A\ll 1$ beyond the dipolar limit \cite{HaradaOptComm1996}. Within such an approach, the time-averaged conservative optical forces acting on the bead directly derive from the potential energies ${\bf F}_{\rm opt} (z,\ell) = -\partial_z U_M (z,\ell)$. The dipolar approach therefore reveals in a straightforward way the crucial property that the optical force field is directly determined from the optical landscape, for every choice of $\ell$. 

Because of the coherent nature of Eq. (1), one dynamical configuration can be selected from the distribution of the successive resonant phase conditions in the $(z,\ell)$ space (corresponding to regions of maximal intensity shown in Fig. 1 (a)). This roots the analogy with a piston-like action exerted by the mirror on the Brownian bead. For instance, picking at $\ell_1$ a resonant phase condition precisely on the waist as shown in Fig. 1 (b) leads to restoring forces that will maintain the bead in a stable trapped position at $z=0$ as displayed in Fig. 1 (c). But a mere compression of the piston to $\ell_2$ brings a bistable configuration where the resonant phase evolution induces regions of local stability from both sides of the waist separated by an unstable point at $z=0$, as seen in the bistable force diagram Fig. 1 (d). 

\begin{figure}[h!]
\includegraphics[width=0.50\textwidth]{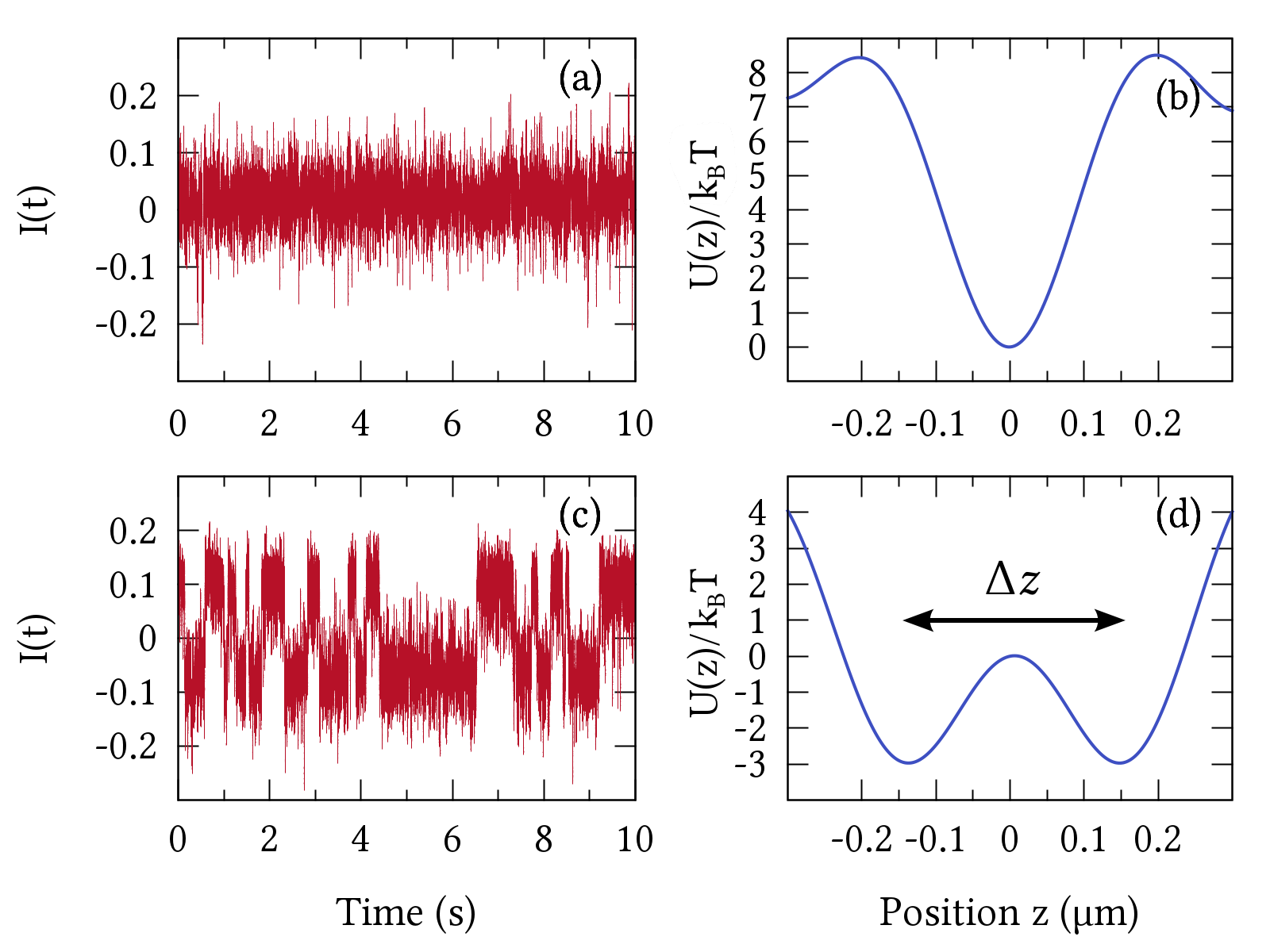}
\caption{Experimental intensity time traces respectively associated with a Brownian motion in a quasi-harmonic well (a) and in a bistable potential between two distant
spatial positions (c). The plots (in units of $k_{\rm B}T$) in (b) and (d) are the potentials associated with the time traces (a) and (c), respectively.
}
 \label{figure_traces}
\end{figure}

The stochastic trajectory $z(t)$ of the bead evolves in every such $\ell$-configuration, modulating the intensity recollected by the objective (see Appendix \ref{AppC}). The time-traces of these modulations allow us to retrieve the essential features of the potentials explored by the bead. For the stable $\ell_1$-configuration of Fig. 1 (c), the time-trace is displayed in Fig. 2 (a) and corresponds to a Brownian motion performed in an quasi-harmonic potential probed by the bead at the bottom of the whole optical potential shown in Fig. 2 (b) -see below. The corresponding trap stiffness is measured by a power spectral density (PSD) analysis \cite{BergSorensenRSI2004}. At low Reynolds numbers, this only relies on the determination of the roll-off trapping frequency and on the knowledge of the fluid friction coefficient $\eta$. Note that we neglect the $\sim 20 \%$ systematic error on the perpendicular viscosity when working at a $2~\mu$m distance from the surface.

The time-trace displayed in Fig. 2 (c) corresponds to the bistable $\ell_2$-configuration of Fig. 1 (d). The intermittency of the intensity signal between the two distinguishable mean values is the signature of the activation of the bead between two metastable positions along the optical axis. There are indeed clearly two different time scales: a short one associated with Brownian fluctuations, and a much longer one on which take place activating events from one to the other of these two positions. In each metastable states, similar time-traces as those of Fig. 2 (a) reveal a quasi-harmonic motion, expected for local equilibrium. Therefore, while the bead performs its Brownian motion within a local well, it diffuses across the potential barrier through rare events thermally assisted \cite{HanggiRMP1990}. 

\begin{table*}[t]
\centering
\begin{tabular}{| l | l | l | l | l | l || l | l | l | l | l |}
  \hline
  $\ell$ ($\mu$m) & M & A & $\Delta{z}$ (nm) & $U_{b1}$ ($k_{\rm b} T$) & $\kappa_b$ (pN/$\mu$m) & $\kappa_1$ (pN/$\mu$m) & $\kappa_2$ (pN/$\mu$m) & $\Delta{U}$ ($k_{\rm b} T$) & $\tau_1$ (s) & $\tau_2$ (s) \\
  \hline
  1.811 & 1.505 & 2.92e-3 & 285 & 2.98 & 2.81 & 3.35 (3.12) & 3.51 (3.85) & -7e-4 (-7e-4) & 0.505 (0.526) & 0.494 (0.474) \\
  \hline
\end{tabular}
\caption{Parameters of the interaction potential extracted from the resolution method applied to the bistable $\ell_2$-configuration. Values directly obtained experimentally are indicated in brackets.}
\label{table1}
\end{table*}

The separation of dynamic time scales and the coherent nature of the optical landscape that provides a built-in spatial reference are two sufficient criteria for applying Kramers' theory to our problem \cite{KramersPhys1940}. In this framework, the interaction potential of the bead is reconstructed for any length of the optical piston, without resorting to any position density probability of the bead along the optical axis that would require an absolute spatial calibration of the setup. As soon as the process is stationary with a sufficient number of recorded activating events, Kramers' theory connects escape rates evaluated from averaged residency time $\tau_{i}$ within each $\{i=1,2\}$ well \cite{VanKampenBook}
\begin{eqnarray}
\frac{1}{\tau_{i}}=\frac{\sqrt{\kappa_{i}}\sqrt{\kappa_b}}{2\pi\eta}\exp\left({-\frac{U_M (z_b,\ell)-U_M (z_{i},\ell)}{k_{\rm B}T}}\right) \label{rates}
\end{eqnarray}
to local trap stiffnesses $\kappa_{i}=\partial^2_z U_M (z,\ell)|_{z_{i}}$ that fix the curvature at the bottom of each well, and to the actual shape of the barrier (position $z_b$ and height) through the absolute value of its curvature $\kappa_b = -\partial^2_z U_M (z,\ell)|_{z_{b}}$. Taking the ratio of both rates therefore leads to measuring the potential energy difference between the local equilibrium positions $\Delta U=U_M (z_2,\ell)-U_M (z_1,\ell)= k_{\rm B}T \ln(\tau_{1}/\tau_{2}\sqrt{\kappa_1 / \kappa_2})$.

Measured $\kappa_{1,2}$ and $\tau_{1,2}$ provide a non-linear system of equations which solution fixes the three ($M$, $\ell$, $A$) parameters needed for a definition of the interaction potential (see Appendix \ref{AppB3}). Experimental values having their own uncertainty, the precision on $\Delta U$ is below $k_{\rm B} T /2$ and below $6$ nm for $\ell$ (see Appendix \ref{AppE}). We also extract from the resolution algorithm the barrier position $z_b$, inverted curvature $\kappa_b$ and height, measured as $U_{b1} = U_{M} (z_b,\ell)-U_{M} (z_1,\ell)$. The barrier, 3 times higher than $k_{\rm B}T$, is still shallow enough to allow the bead mapping, through thermal fluctuations, the bistable potential around $z=0$. The distance $\Delta z$ over which the bead is activated is also measured. From the parameter values gathered in Table I, the interaction potential profile can be plotted as a function of the bead displacement as in Fig. 2 (d) in units of $k_{\rm B}T$. We stress that the phase structure of ${I}^{\rm opt}_{M}$ forbids a simple $4^{\rm th}$-order potential (i.e. Duffing type).

Further insight comes from looking at the system from the point of view of the evolution of configurations controlled by the external variable $\ell$. An incremental change ${\rm d}\ell$ of the length of the piston pushes the bead out of equilibrium and forces it to relax in the new $\ell + {\rm d}\ell$ configuration with a stiffness $\kappa$ and a fluid friction $\eta$ on a time $t_D\sim \kappa /\eta$ set by diffusion, typically ca. $10^{-2}~{\rm s}$ in our conditions. Hydrodynamic effects on the bead due to the motion of the mirror can be neglected since the incremental shift of the mirror by $|{\rm d}\ell|=20$ nm, performed with a speed of $1~{\rm mm}/{\rm s}$ set for the piezo-actuator, is associated with a low $10^{-8}$ Reynolds number. Accordingly, the displacement of the fluid remains purely diffusive and the moving piston therefore has no direct mechanical action on the bead. Under such conditions, the only source of mechanical loss in the system is given by the relaxation process from one configuration to the other. 

We emphasize that $z(t)$ does not map the entire canonical equilibrium distribution associated with an $\ell$-configuration. It only maps a thermally accessible subset of it, that can be resolved for sufficiently long acquisition times ($30$ s in our experiment). The notion of stability then corresponds to local stable wells much deeper than $k_{\rm B}T$ while bistability corresponds to local barrier heights of the order of $k_{\rm B}T$ over which the bead can be activated. In this picture, stable phases can be identified from bistable phases, as drawn in Fig. 3 (a) in the $M-\ell$ parameter space. 

Our coherent optical piston configuration gives a unique capacity in monitoring the crossover between these phases. Indeed, a continuous compression of the piston connecting two stable configurations forces the bead to go through a whole phase of bistability, starting for a piston length $\ell_i$ with the bead in an initial stable position at the incident waist $z=0$ and ending for $\ell_f<\ell_i$ with the bead in the very same spatial position but within a different stable potential. From our resolution method, the bistable dynamics of the bead can be solved for each step in $\ell$. The ($M$, $\ell$, $A$) parameters fixing each potential throughout the bistable phase are extracted and the actual steps of the entire path followed by the system between $\ell_i$ and $\ell_f$ can be plotted in the $M-\ell$ plane. It is worth noting that the resolved $\ell$ values follow precisely the mirror actuation command and that the path shows a small dispersion in $M$ values. This important outcome of the analysis leads to determine the potential profiles even in the stable phase from a mere extrapolation on the variable $\ell$. This is done for instance for the stable $\ell_1$-configuration of Fig. 1 with the measured potential profile plotted in Fig. 2 (b).

As shown in Fig. 3 (b), the path can also be represented through intensity probability densities, associating to each probability density extrema a well of the resolved potentials. These plots clearly reveal the progressive onset of a bistable dynamics of the bead along the optical axis of the setup. As a clear advantage of our statistical method, this crossover dynamics can be probed despite the unknown exact relation between measured intensities and bead positions along the optical axis.

It is possible to give a thermodynamic description of the path from an incremental energy balance. Following \cite{SekiBook}, this can be drawn from the Langevin over-damped dynamics of the bead accounting for the contribution of the piston with
\begin{eqnarray}
{\rm d}U_M(z,\ell)={\rm d}Q+\partial_{\ell} U_M(z,\ell){\rm d}\ell ,
\end{eqnarray}
where ${\rm d}U_M(z,\ell)$ is the change of the potential energy of the bead. This change comes from two sources: {\it (i)} the heat balance ${\rm d}Q=\xi{\rm d}z-\eta\dot{z}{\rm d}z$ between fluctuation (determined by a thermal stochastic force $\xi$) and friction (related to $\eta$), and {\it (ii)} the external work done on the bead by the displacement of the piston where the external variable $\ell$ controls the evolution of the potential configurations. 

By waiting much longer than $t_D$ between each incremental change ${\rm d}\ell$, steady-state of every new configuration is reached through mechanical relaxation of the bead. This insures that the system evolves through the configurations in an iso-thermal and quasi-static way. Moreover, all incremental changes in the optical potential are kept smaller than $k_{\rm B}T$. This implies that the bead would go back exploring the same configurations if the displacement of the piston would be reversed. As a consequence, the whole path in Fig. 3 is thermodynamically reversible. 

Under such conditions, the averaged external work is directly related to the Helmholtz free energy with $\langle \partial_{\ell} U_M(z,\ell)\rangle_{\{z\}} {\rm d}\ell={\rm d}F(\ell)$ \cite{SekimotoProgTheo1998,BlickleNatPhys2012}. The averaging process being performed over the positions occupied by the bead in the given configuration, the free energy is only function of the external variable $\ell$. The total amount of work performed by the piston through the isothermal reversible process is directly given by the free energy difference between the initial $\ell_i$ and final $\ell_f$ positions of the piston which can be calculated from the $(\ell_i,\ell_f)$ canonical partition functions as 
\begin{eqnarray}
W=F(\ell_f)-F(\ell_i)=-k_{\rm B}T\ln\left[ \frac{Z(\ell_f)}{Z(\ell_i)}\right].
\end{eqnarray}

For both initial and final stable configurations, the partition functions can be expanded to second order around the waist as
\begin{eqnarray}
Z(\ell_{i,f})\simeq e^{-\frac{U_M(0,\ell_{i,f})}{k_{\rm B}T}}\cdot \sqrt{\frac{2\pi k_{\rm B}T}{\kappa_{i,f}} } 
\end{eqnarray}
where $\kappa_{i,f}=\partial^2_z U_M(z,\ell_{i,f})|_{z\sim 0}$ are the stiffnesses of the $(\ell_i,\ell_f)$ stable potentials. 
\begin{figure}[htb!]
\includegraphics[width=0.50\textwidth]{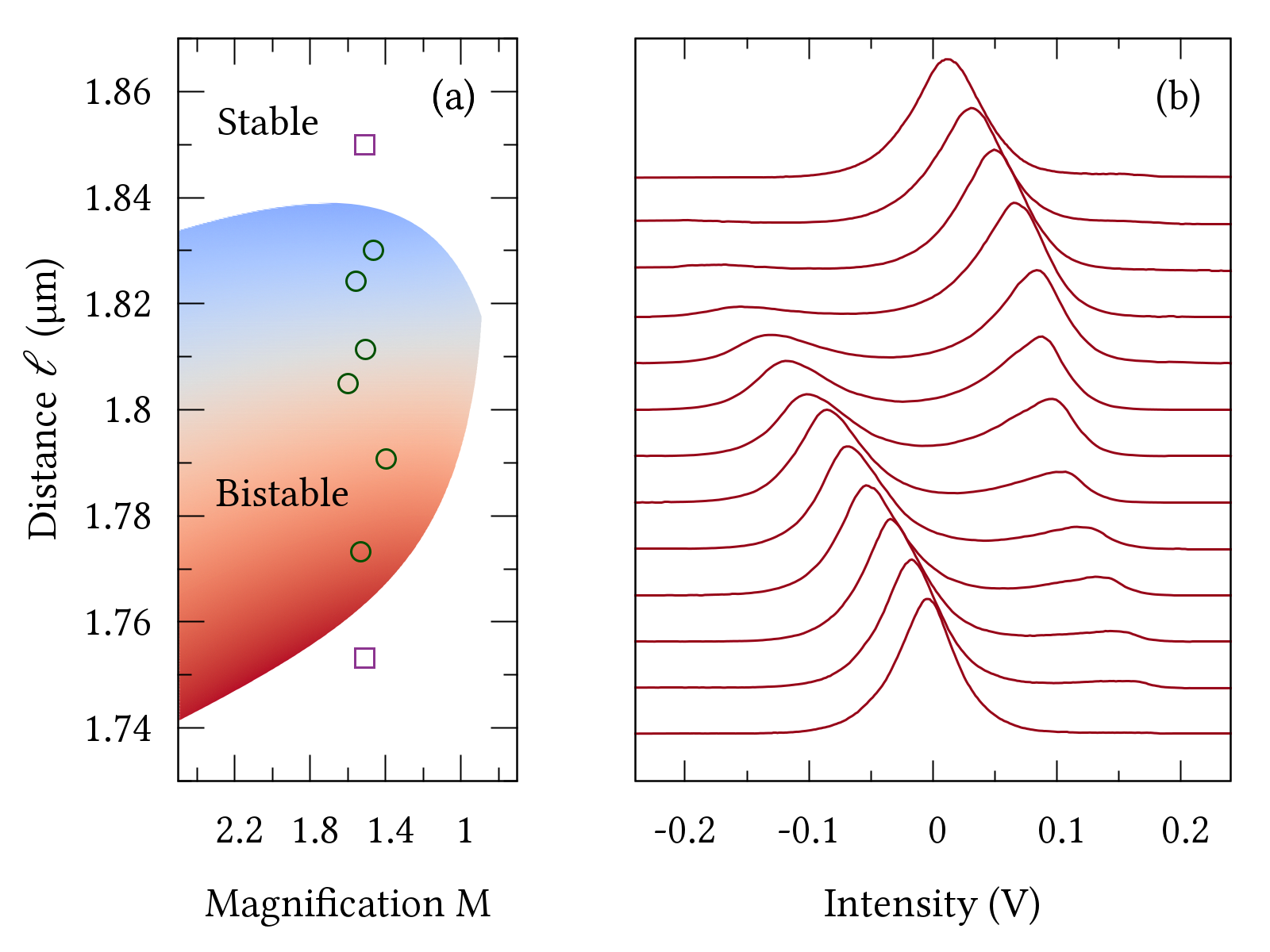}
 \caption{(a) Regions of stability and bistability in the $\ell-M$ parameter space. Experimentally solved bistable configurations are displayed in open circles as $\ell$ is varied. The points in the stable region (open squares) are extrapolated from the extreme bistable points according to the $20$ nm piezo-actuation. The color map in the bistable phase codes the asymmetry of the bistable potential as $\Delta{U} / (U_M(z_1,\ell) + U_M(z_2,\ell))$ positive in red, negative in blue. (b) Intensity probability densities (for $30~{\rm s}$ acquisition) as the mirror distance $\ell$ is reduced, crossing over the bistability region between two stable bead dynamics. The six central bistable plots correspond to the six open circles plotted in (a). 
 }
 \label{fig_transition}
\end{figure}
The total energy balance $\Delta U_M=W+Q_{\rm rev}$,
with 
\begin{eqnarray}
&&\Delta U_M = U_M(0,\ell_{f})-U_M(0,\ell_{i})   \nonumber  \\
&&Q_{\rm rev} = k_{\rm B}T\ln \left[\sqrt{ \frac{\kappa_i}{\kappa_f}}\right],
\end{eqnarray}
connects, along the path, the total amount of energy change $\Delta U_M$ to the heat $Q_{\rm rev}$ produced by the whole reversible process. For the crossover of Fig. 3, $\Delta U_M=-2.34\times 10^{-21}~{\rm J}~ (\pm 3\%)$. As clearly seen, a stiffness difference between the initial $\ell_i$ and final $\ell_f$ configurations is directly related to the production of heat. We unambiguously calculate a quantity of reversible heat $Q_{\rm rev}=-2.48\times 10^{-22}~{\rm J}~ (\pm 10\%)$ transferred to the fluid by the bead along the path (see Appendix \ref{AppE} for the evaluation of the uncertainties). The negative $Q_{\rm rev}$ value means that friction dominates over fluctuation as the source of heat. This is consistent with the fact that the bead is displaced from an initial stable $\ell_i$-configuration to a final $\ell_f$ one which is optically more confined. The small $Q_{\rm rev}$ value quantifies the deviation from adiabaticity with $W>\Delta U_M$. This deviation stems from the mechanical deformation of the interaction potential at both ends of the path which is due to an increase in the optical intensity as the mirror gets closer to the waist. 

In essence, our optical piston configuration enables to control the source of heat. This could lead to the possibility to reach adiabaticity with $W=\Delta U_M$ in a simple way. For instance, in a pure standing wave configuration, both ends of the path have identical trapping stiffnesses leading to $Q_{\rm rev}=0$. In this context, tailoring the optical landscape is particularly appealing. It leads to the possibility to induce and probe all kinds of thermodynamic processes through the control of both heat production and potential energy differences. Because these quantities are optically determined, the level of control available is expected to be much smaller than $k_{\rm B}T$.

\section*{Acknowledgments}

We thank A. Canaguier-Durand, A. Cuche, J.A. Hutchison, T.W. Ebbesen and S. Reynaud for fruitful discussions and support. This work was funded in part by the ERC (grant 227577) and the ANR (Equipex ``Union'').

\appendix

\section{Experimental Setup}  \label{AppA}

The optical setup is sketched in Fig.\ref{fig.exp.scheme}. A linearly polarized TEM$_{00}$ beam from a CW diode-laser (Excelsior Spectra-Physics, wavelength $\lambda=785~{\rm nm}$, power $45$ mW) is sent into a dry objective (Nikon CFI Plan Fluor 60X, 0.85 NA) and focused in a water cell (deionized water, $80~\mu$m thick) enclosing mono-dispersed dielectric polystyrene beads (Thermoscientific Fluoro-Max Red Dyed, refractive index $1.58$) of radius $500$ nm. The cell is topped by a $170~\mu$m thick cover slip. Spherical aberrations are compensated by the objective (correction ring set to $0.2$ mm).

The laser beam traps a single bead in the vicinity of a movable mirror ($300$ nm thick evaporated gold film on a glass substrate). The beam is reflected by the mirror and recollected by the objective. It is sent to a non-polarizing cube beamsplitter where it is equally divided. One arm of the intensity signal is vignetted by a pin-hole and recorded by a PIN photodiode (Thorlabs Det10A). Amplified before numerical conversion and acquisition (NI PCI-6251, 16 bits resolution), this port provides intensity time-traces that measure the axial displacement $z(t)$ of the bead inside the optical trap. The second port is sent to a CCD camera (Allied Guppy Pro F-031) that images the recollected beam spot.

We took care to isolate optically the CW diode-laser, using a free-space Faraday isolator (Thorlabs IO-5-NIR-LP). The isolation is further improved by injecting the laser beam into the objective using a polarizing cube beamsplitter coupled to a quarter-wave plate. This prevents as much as possible the recollected signal to be send back to the injection port. This scheme implies that the optical landscape created between the objective lens and the mirror is the coherent superposition of a forward right-handed circularly polarized beam and a backward left-handed circularly polarized beam. 

\begin{figure}[!htb]
  \centering
  \includegraphics[width=0.45\textwidth]{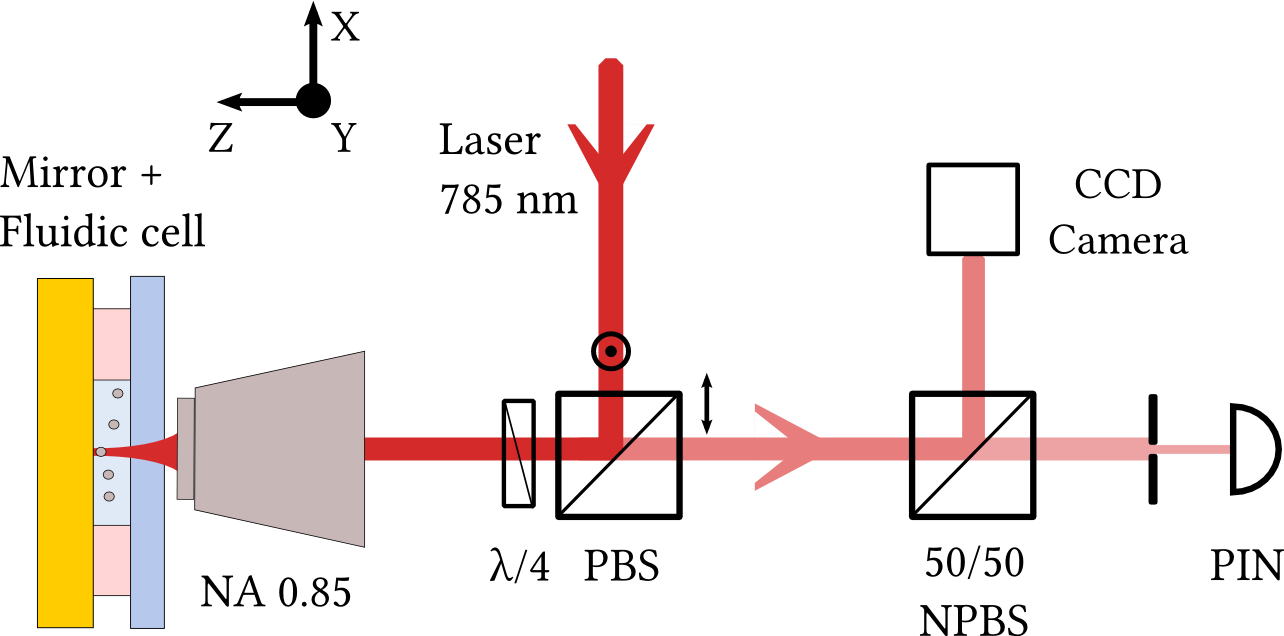}
  \caption{Schematics of the experimental setup, indicating in particular the secondary isolation stage at the level of the polarizing beam splitter (PBS) and the added quater-wave plate $\lambda / 4$. The two ports from the non-polarizing beam splitter (NPBS) used for recording the recollected beam are also represented. In the chosen frame, a right-handed circularly polarized field propagating along the $z>0$ direction is described with $\boldsymbol{\sigma}_+ = (\hat{\bf x}-i\hat{\bf y})/\sqrt{2}$.}
  \label{fig.exp.scheme}
\end{figure}

\section{Solving the interaction potential} \label{AppB}

\subsection{Optical Fields}  \label{AppB1}

The optical fields ${\bf E}_{\rm tot}$ created inside the optical piston (i.e. between the objective lens and the mirror) is given by the coherent superposition of an incident and a reflected Gaussian beams. The incident beam is, as explained above, right-handed ${\boldsymbol \sigma}_{+}$ circularly polarized and described by its Rayleigh range $z_{R}$ and its waist $w_0$ position fixed at $z = 0$
\begin{align}
  {\bf E}^{+k}(z) = E_{00} \frac{w_{0}}{w(z;z_R)} \exp\left(i k z - i \xi(z;z_R)\right){\boldsymbol \sigma}_{+}
\end{align}
with
\begin{align}
w(z;z_R) &= w_{0} \sqrt{1 + \left(\frac{z}{z_{R}}\right)^{2}} \\
\xi(z;z_R) &= \mathrm{arctan}\left(\frac{z}{z_{R}} \right)
\end{align}
and $ z_{R} = {\pi w_{0}^{2}}/{\lambda} $, $ n_{2} \lambda = \lambda_{0} $, the optical wavelength in water (refractive index $n_2$) and $\pi w_{0} = {\lambda}/{ \mathrm{NA}}$. 

The reflected field, counter-propagating $k\to -k$ with respect to the incident beam, is multiplied by a reflectivity coefficient $ r(\lambda) $. Its waist position is the mirror-image of the incident waist position. But before being reflected, the incident beam is intercepted by the bead because the bead diameter is larger than the incident waist $w_0$. Considering that the refractive index of the bead (polystyrene) is different from that of the fluid (water), the incident beam transmitted through the bead is magnified, the bead considered to act as a lens-doublet. We account of this effect by introducing an effective magnification parameter $M$ on the reflected beam itself. This leads to changing the Rayleigh range $z_R \to M^2 z_R$ and waist $ w_{0} \to M w_{0}$ of the reflected beam with respect to the incident beam. The reflected beam, left-handed ${\boldsymbol \sigma}_{-}$ circularly polarized, is then expressed as:
\begin{align}
  {\bf E}^{-k}_{M}(z - 2\ell) = & E_{00} \frac{M w_{0}}{w(z;M^2 z_R)} \times \nonumber \\
  &\hspace{-20mm} \exp\left(-i k (z-2\ell) + i \xi(z-2\ell;M^2 z_R)\right){\boldsymbol \sigma}_{-}.
\end{align}

The coherent superposition of the incident and the magnified reflected Gaussian beams determines the optical landscape of the problem. The corresponding optical intensity writes as
\begin{align}
\begin{split}
&I^{{\rm opt}}_{M}(z,\ell) = |{\bf E}^{+k}(z) + r[\lambda]\cdot {\bf E}^{-k}_{M}(z - 2\ell) |^2  \\
  & = E_{00}^{2} \frac{w_{0}^{2}} {w^{2}(z;z_R)} + \rho^{2} E_{00}^{2} \frac{w_{0}^{2}} {w^{2}(z-2\ell;M^2 z_R)} \\
  & + \frac{2 \rho E_{00}^{2} M w_{0}^{2}} {w(z;z_R) w(z-2\ell;M^2 z_R)} \times \\
 & \hspace{15mm} \cos \left(2 k (z - \ell) - \xi(z;z_R) - \right.  \\
  &\hspace{25mm}\left. \xi(z-2\ell;M^2 z_R) + \psi\right).
\end{split}
\end{align}
with an interference term between the two beams. In the vicinity of the waist, it shows that the modulations of the optical landscape (that will eventually correspond to the local potential barriers, as discussed below) are determined from an harmonic term. This immediately stresses that a standard $4^{\rm th}$ order polynomial description of the barrier -as done when resorting to a typical Duffing model- is not appropriate for our optical piston configuration. 

\subsection{Conservative optical force} \label{AppB2}

In the dipolar regime, the bead is characterized by an electric polarizability $\alpha$. Neglecting any source of dissipation within the bead, i.e. assuming that ${\rm Im}[\alpha ]\sim0$, the gradient force is the only force exerted on the bead
\begin{equation}
{\bf F}_{\rm {opt}} = - \partial_z U_M(z,\ell).
\end{equation}
It directly derives from the interaction potential energy determined from the optical landscape intensity as
\begin{equation}
U_{M}(z,\ell) = - A \cdot \frac{\alpha}{2 \epsilon_{0} n_{2} c} I_{opt}(d,z)
\end{equation}
with $\alpha = 4 \pi \epsilon_{0} n_{2}^{2} a^{3} [(n_{1}/n_{2})^{2} - 1] /  [(n_{1}/n_{2})^{2} + 2]$, $n_{1}$ the refractive index of the bead,
$n_{2}$ the refractive index of the fluid, $a$ the radius of the bead and $\epsilon_0$ the vacuum permittivity.

Despite the fact that is is related to the bead geometry, the $M$ parameter value for the magnification by the bead is not necessarily a constant of the model as it depends on the width of the beam intercepted by the bead. The $A$ parameter quantitatively corrects the value of the potential calculated in our Rayleigh-based model due the finite size of the bead that should be accounted for in a more realistic description of the optical interaction. In fact, as given in the main text, the interaction potential turns out to be smaller by approximately 3 orders of magnitude for a $1~\mu$m bead. This value is in good agreement with existing evaluations for the size effect \cite{HaradaOptComm1996}. We stress that because $A$ is a parameter of the model that characterizes the intensity of the coupling of the bead with the optical intensity, it is kept fixed once determined for a given bead.

\subsection{System Resolution}  \label{AppB3}

Monitoring the instantaneous motion $I(z(t))$ of the bead in the bistable phase, mean residency times $\tau_{1,2}$ and stiffnesses $\kappa_{1,2}$ are extracted from experimental intensity time traces for each well. Kramers rate equations provide the following system fed with the extracted values:
\begin{align}[left = \empheqlbrace]
\partial^2_z U_M (z,\ell)|_{z=z_{1}} &= \kappa_1 \\
\partial^2_z U_M (z,\ell)|_{z=z_{2}} &= \kappa_2 \\
\frac{\sqrt{\kappa_{1}}\sqrt{\kappa_b}}{2\pi\eta}\exp\left({-\frac{U_M (z_b,\ell)-U_M (z_{1},\ell)} {k_{\rm B}T}}\right) &= \frac{1}{\tau_1} \label{S3} \\
\frac{\sqrt{\kappa_{2}}\sqrt{\kappa_b}}{2\pi\eta}\exp\left({-\frac{U_M (z_b,\ell)-U_M (z_{2},\ell)} {k_{\rm B}T}}\right) &= \frac{1}{\tau_2} \label{S4}
\end{align}
where Eqs. (\ref{S3}) and (\ref{S4}) are escape equations given by Kramers theory in the over-damped regime for each of the metastable states of the bistable phase.
Resolution of this system gives access to the remaining unknown quantities of our model: the mirror waist distance $\ell$, the bead magnification effect $M$ and the size correction parameter $A$ related to the coupling between the light field and the finite size bead.

The energy difference between the two metastable wells can be derived from the extracted residency times and stiffnesses as $\Delta{U} = k_{\rm B}T \ln\left({\frac{\tau_{1}}{\tau_{2}}\sqrt{\frac{\kappa_1}{\kappa_2}}}\right)$. Taking the ratio of equations \ref{S3} and \ref{S4} then removes the dependency of the system on the properties of the barrier but requires the knowledge of $A$. Fixing $A$ and using the simplex algorithm, the external variable $\ell$ (piston length) and the magnification parameter $M$ are determined. This therefore gives the entire potential $U_M (z,\ell)$ for a given $A$. Iterating this resolution over $A$ until the rate equations are verified provides the triplet ($\ell, M, A$) that best solves the whole system. Once determined in one $\ell$-configuration (say the symmetric bistable configuration), the parameter $A$ is then kept constant when solving other bistable configurations (for instance varying $\ell$).

\section{Axial Displacement} \label{AppC}

The bistable behavior of the bead can be monitored on the CCD camera because the motion of the bead along the optical axis and across the bistability barrier changes the Gaussian envelope of the reflected beam and thus the diffraction pattern of the recollected beam imaged on the camera, as shown in Fig (\ref{fig.exp.bishot}). Thus, by the sole measurement of the recollected intensity, one can access part of the bead dynamics. Nevertheless, the low acquisition rate of the CCD camera is a strong limitation for analyzing precisely the stochastic motion of the bead.

\begin{figure}[!htb]
  \centering
  \includegraphics[width=0.23\textwidth]{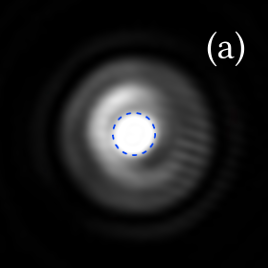}
  \includegraphics[width=0.23\textwidth]{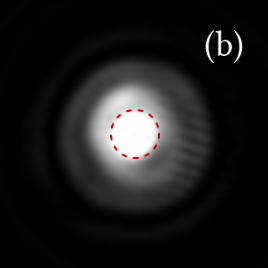}
  \caption{Recollected spots imaged on the CCD camera in the case of a bead trapped in a bistable configuration. with same exposition time. Panel (a) shows the bead in its most distance position from the mirror, i.e. the position behind the waist (z $<$ 0). Panel (b) shows the bead in front of the waist, thus closer to the mirror. Both images have been recorded with the same exposure time. The recollected intensity is higher near the mirror (image (b)) than away from it. The central area of the recollected spot is indicated by the superimposed circles, with a 44 pixel-diameter for (a) and 50 for (b).  }
  \label{fig.exp.bishot}
\end{figure}

To do so, we rather measure the recollected intensity using a PIN photodiode that grants a better sensitivity and a high acquisition rate. We set this rate for our experiments at $2^{18}= 262144$ Hz using low noise preamplifiers (SR560). The PIN signal for intensity measurements was recorded in AC mode thus filtered through a $0.3$ Hz high-pass filter at $6$ dB/oct to remove the continuous component of the signal. High-pass filtering poses no issue since we focus on signal fluctuations while it allows, after amplification, to span the signal of interest over the whole acquisition card input range. For the slow varying signals, the exponential decrease coming from the high-pass filter is later compensated numerically to within a constant. A low-pass filter at $100$ kHz at $6$ dB/oct was also used to avoid aliasing.

Time traces measured by the PIN photodiode clearly reveal the bistable dynamics of the bead as shown in Fig \ref{plot.timetrace}. The intensity time trace is distributed around two mean values from which we extract and concatenate the dynamics associated with each well of the bistable potential, as shown on Fig \ref{plot.timetracesplit}. The full signal dynamics clearly appears stationary: the signal stays centered around a constant value (near 0) and the ratios between the mean residency times for each mean value is time-independent. Similarly, the two concatenated time traces reveal stationarity properties.

\begin{figure}[htbp]
  \centering
  \includegraphics[width=0.5\textwidth]{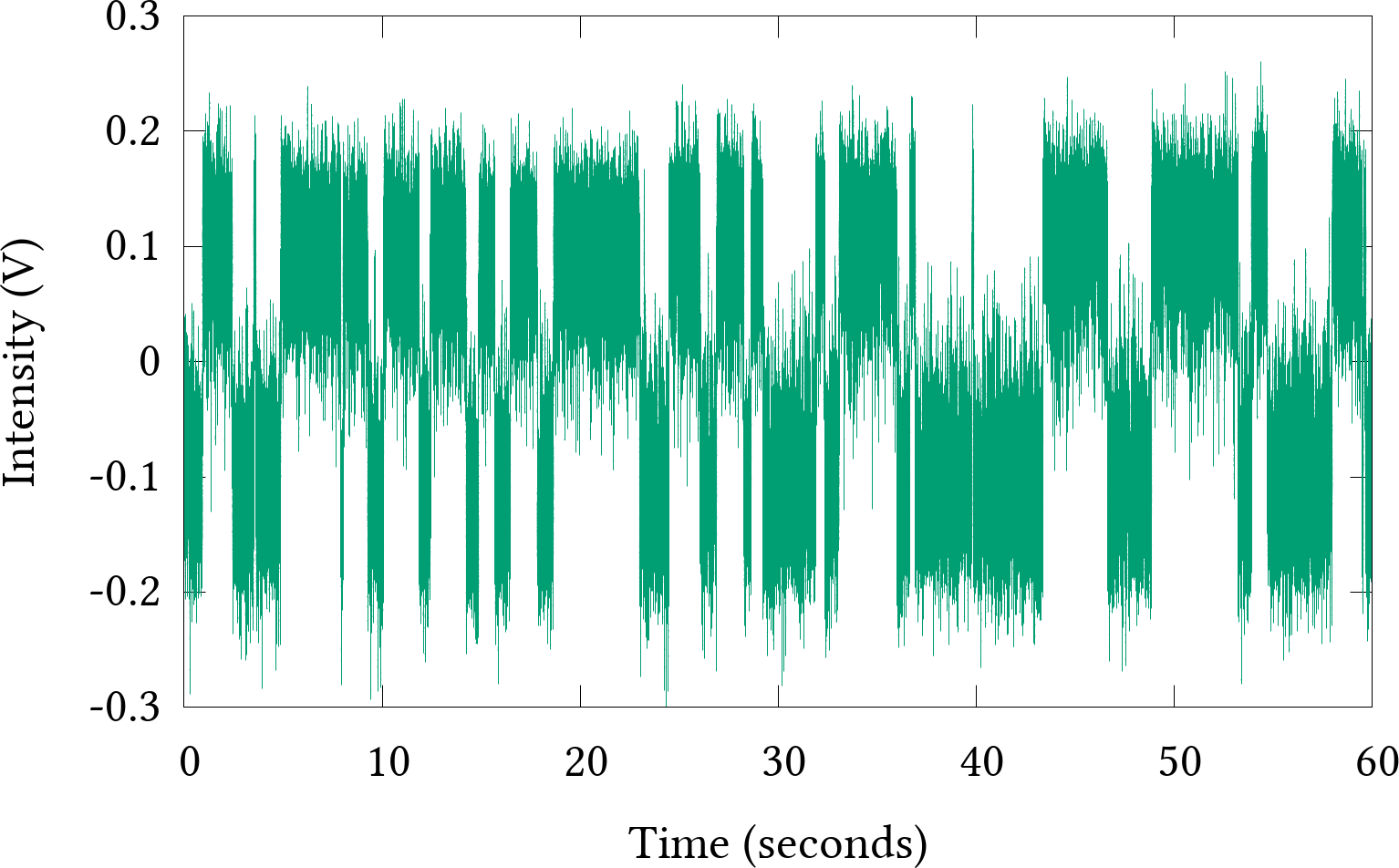}
  \caption{Time trace associated with a bistable motion of the bead as recorded by the PIN photodiode on the reflected beam intensity. The acquisition time is set to $60$ s, at a rate of $262$ kHz.
  }
  \label{plot.timetrace}
\end{figure}
\begin{figure}[htbp]
  \centering
  \includegraphics[width=0.5\textwidth]{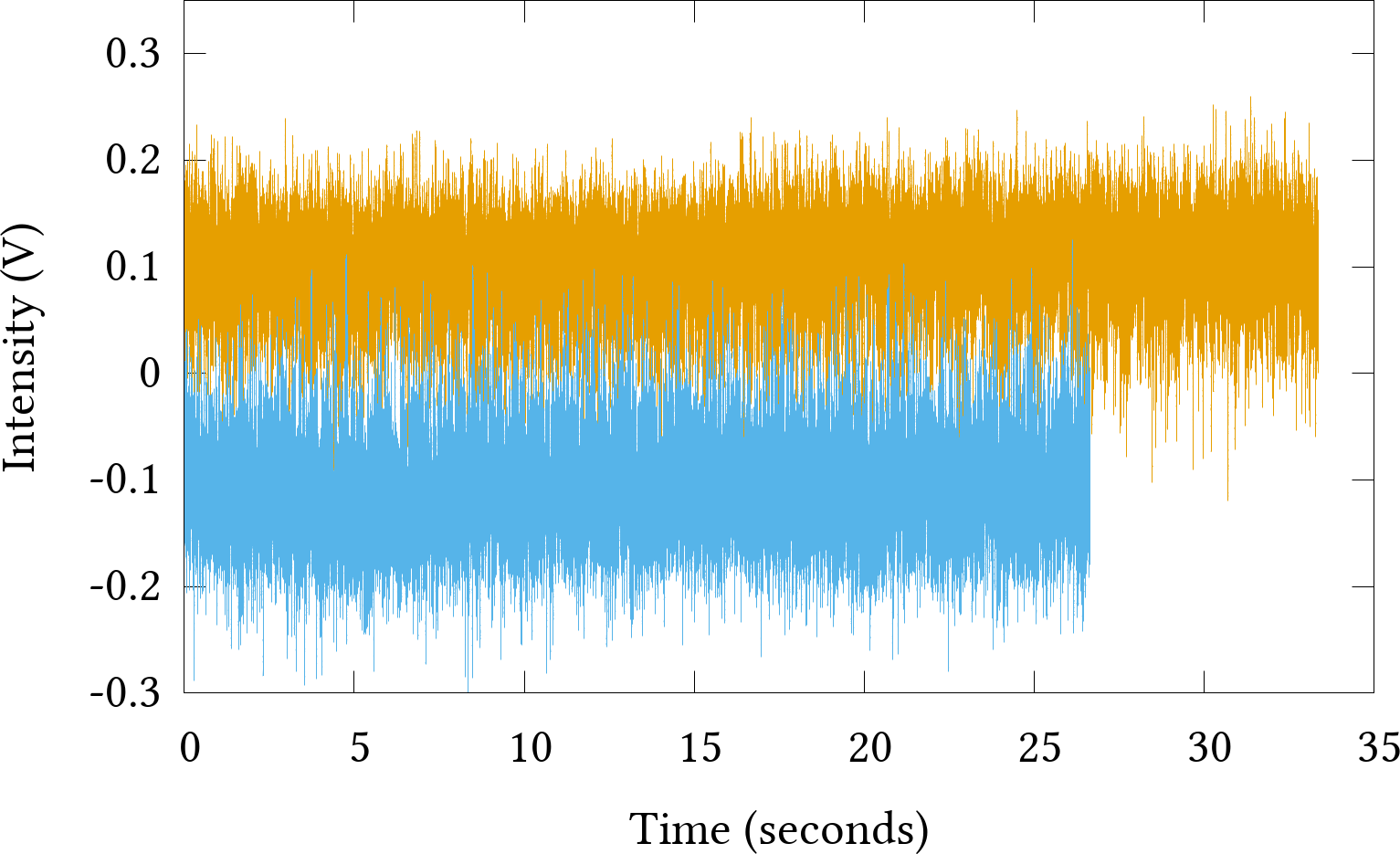}
  \caption{Concatenated time traces associated with each of the mean values measured in the time trace of Fig. \ref{plot.timetrace}.}
  \label{plot.timetracesplit}
\end{figure}

In the experiment, we spatially filtered the recollected beam before the PIN photodiode through a pinhole (typical aperture of about $1$ mm$^2$). This filtering was useful since it actually enhanced the separation between the two average intensities in the bistable configurations. 

\section{Power Spectral Density}   \label{AppD}

\begin{figure}[!htbp]
  \centering
  \includegraphics[width=0.45\textwidth]{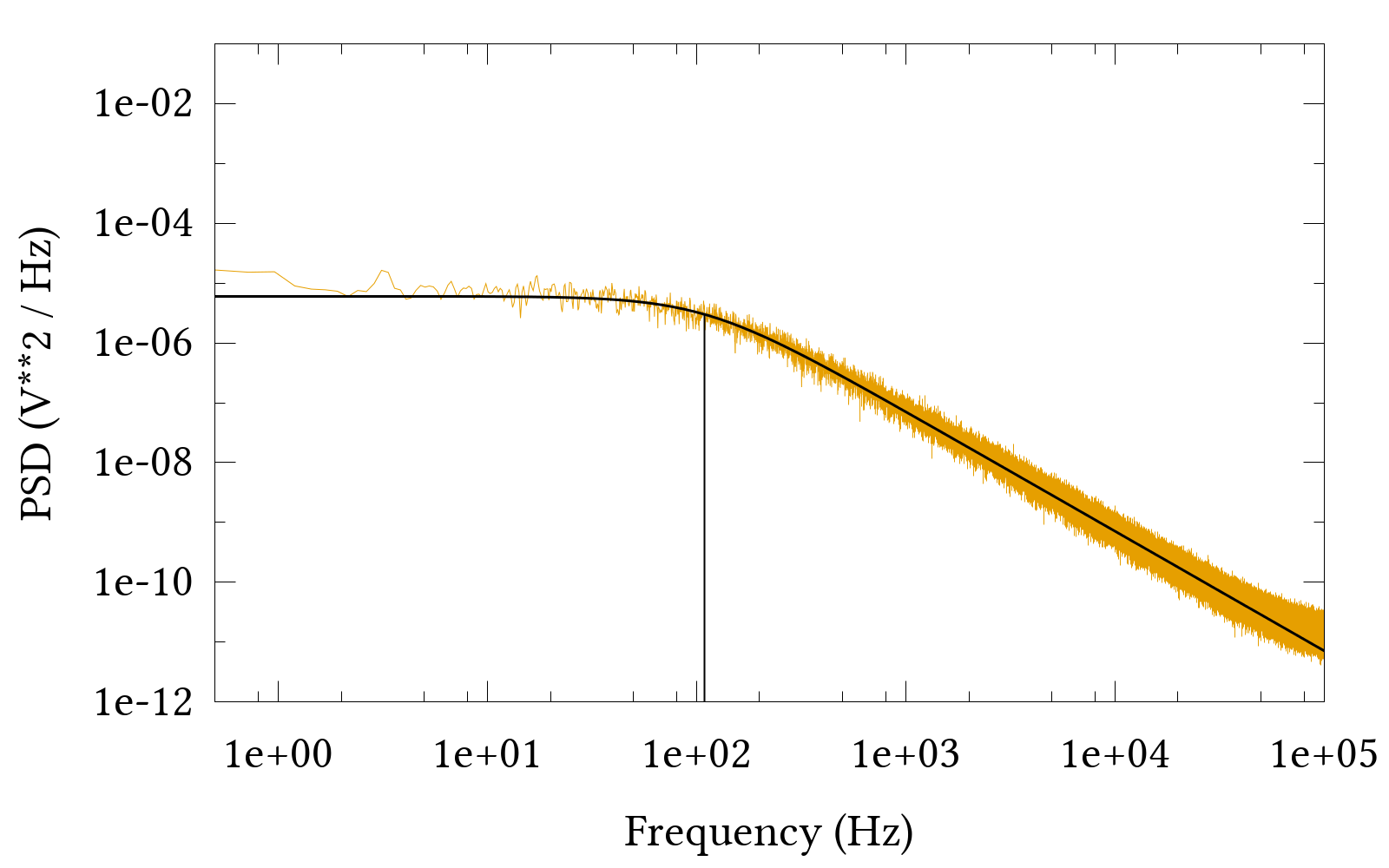} \\
  \includegraphics[width=0.45\textwidth]{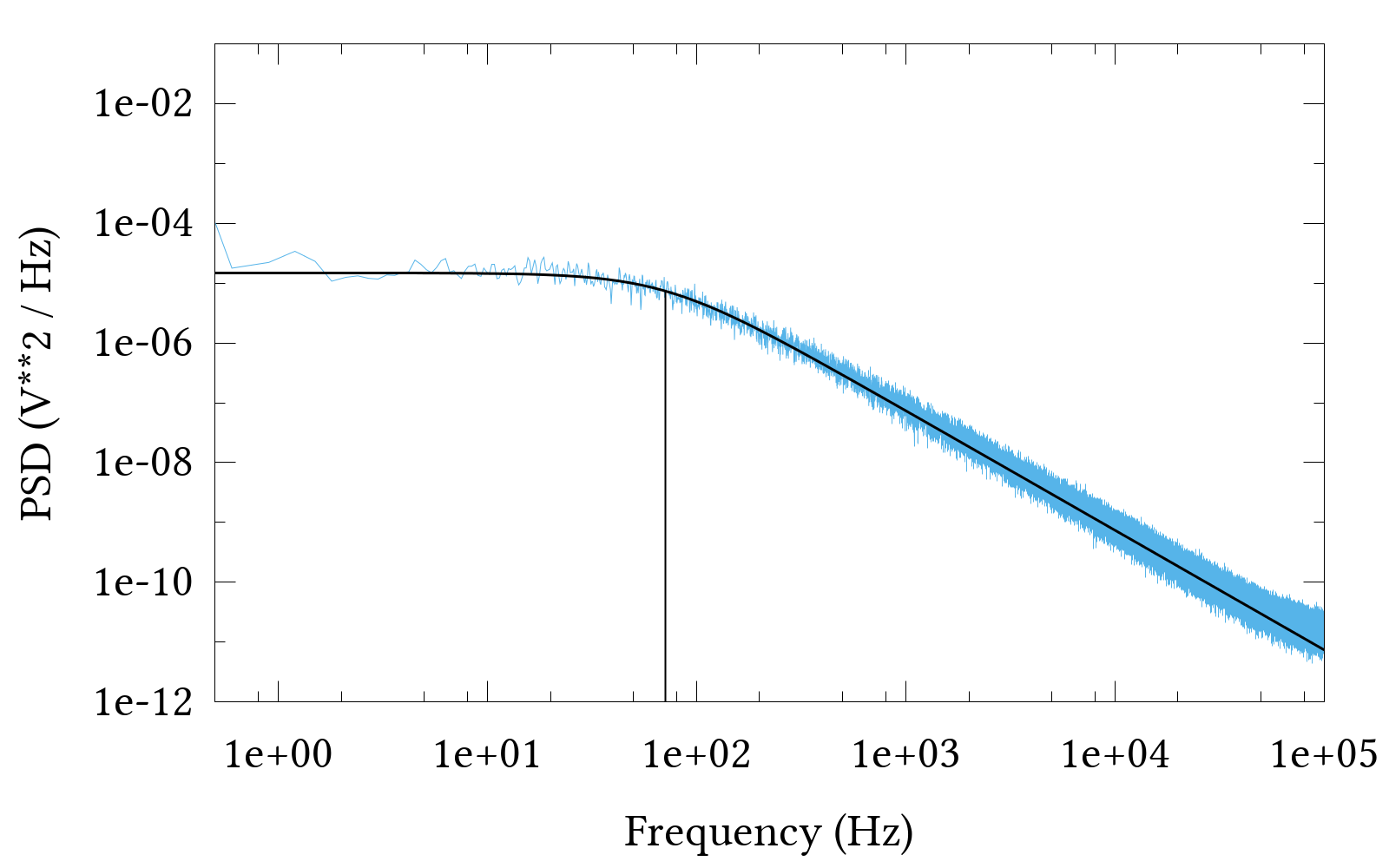}
  \caption{PSD taken on each of the concatenated time traces given in Fig. \ref{plot.timetracesplit}. Upper curve is the PSD for the $I(t) > 0$ concatenated time trace and the bottom one the PSD of the concatenated time trace when the bead is in the well further away from the mirror (corresponding to $I(t) < 0$). The black solid line is the Lorentzian fit of the data and the vertical line gives the roll-off frequency of each of the quasi-harmonic trap associated with each locally stable positions.}
  \label{plot.PSDs}
\end{figure}

\begin{figure}[!htbp]
  \centering
  \includegraphics[width=0.5\textwidth]{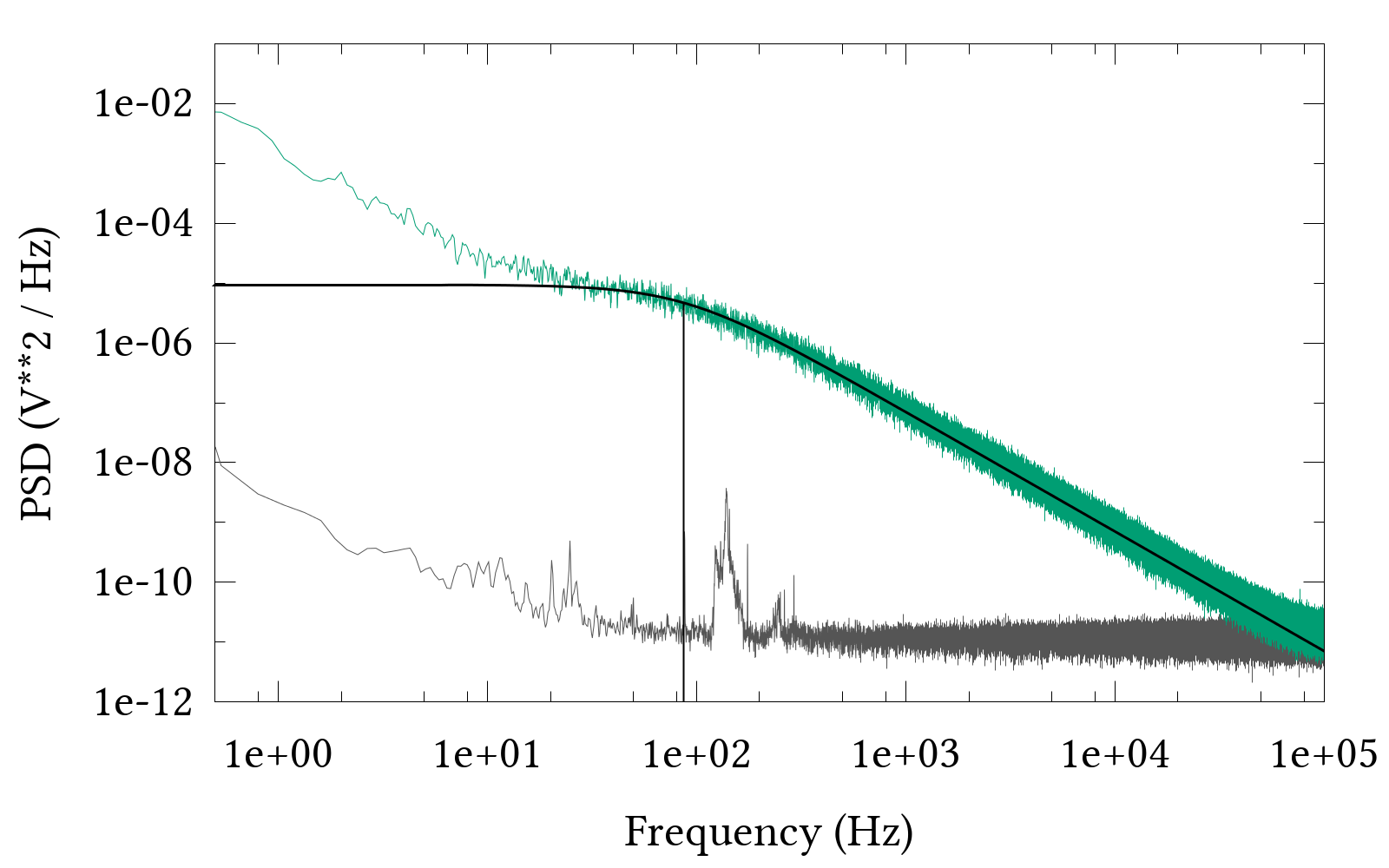}
  \caption{PSD taken directly from the full time trace shown in Fig. \ref{plot.timetrace} for an acquisition time of a minute. The bottom PSD (dark)  corresponds to the experimental laser noise.}
  \label{plot.PSDfull}
\end{figure}

Assuming that the recorded stationary intensity time traces are ergodic, meaning that a time trace is independent of the initial position of the bead and leads to the same distribution for different realizations with identical parameters, one can study the bead dynamics through its power spectral density (PSD) \cite{Webb1}. This standard approach has the advantage of being straightforward to apply on rapidly fluctuating data.  

It clearly appears that the PSDs associated with each of the concatenated time traces of Fig. \ref{plot.timetracesplit} follow the typical Lorentzian shape of a Brownian motion performed in an harmonic trap in the over-damped regime, as seen in Fig. \ref{plot.PSDs}. This implies that the two wells of the bistable potential, separated by the activation barrier, are quasi-harmonic, with stiffnesses $\kappa_{1,2} = 2 \pi\, \eta\ f_{1,2}$ that can be determined directly from the fluid drag $\eta$ and the so-called roll-off frequency $f_{1,2}$ of the trap measured on each PSD \cite{BergSorensenRSI2004}. In the full signal PSD shown in Fig. \ref{plot.PSDfull}, these quasi-harmonic traps are seen through the Lorentzian fit at high-frequencies. But the spectrum within these local wells does not exhaust the bistable dynamics. Low frequencies indeed reveal a strong increase in the power spectrum which is due to the activation process of the bead over the bistable barrier, occurring on a typical $\sim 1$ Hz regime. In other words, the crucial separation of time scales discussed in the main text is readily observed on the PSD associated with the bistable motion of the bead.

\section{Experimental Uncertainties}  \label{AppE}

We assume that the distribution of the measured residency times are Poissonian, implying that their mean values $\tau_{1,2}$ equal their variances $\sigma_{\tau_{1,2}}$. Measuring $N=25$ number of back and forth activations of the bead through the barrier (over an acquisition time of $T=30$ s) for the bistable configurations described in the main text, leads to an experimental uncertainty in the $\tau_{1,2}$ determination of $\delta\tau_{1,2} = \sigma_{\tau_{1,2}} / \sqrt{N} =  0.2 \ \sigma_{\tau_{1,2}}$. Because the signal is stationary, $\tau_{1} + \tau_{2} = \frac{T}{N}$ and therefore $\tau_{2} = \frac{T}{N} - \tau_{1}$.

In addition, by taking the expression of the perpendicular viscosity \cite{Padgett2009} and its derivative, and neglecting the systematic error in the region where the bead evolves, the uncertainty in the change of viscosity along the displacement of the bead can be estimated at a $3\%$ level over $300$ nm bed displacement. To this uncertainty, $1~\%$ is added from the extraction of the roll-off frequency at the level of the PSD (fit uncertainty). This thus leads to a global stiffness uncertainty of about $\delta\kappa_{1,2} = 0.04 \ \kappa_{1,2}$.

The $6$ resolved $M$ values have mean and sample deviation respectively of $\overline{M} = 1.508$ and $\sigma_M = 0.071$. The uncertainty therefore is $\delta{M} = { \sigma_M} / {\sqrt{6}} = 0.029$ and $M = \overline{M} \pm \delta{M} = 1.508 \pm 0.029$ which is an uncertainty of $4\%$ of the value.

These uncertainties are propagated \cite{JRTaylor} to determine the uncertainty on $\Delta{U}$ used as an input (through  $\kappa_{1,2}$ and $\tau_{1,2}$) in our system solver. Propagations of $\delta\tau_{1}$ and $\delta\kappa_{1,2}$ in $\Delta{U}$ in bistable configurations where $\tau_{1} \sim \tau_{2}$ lead to
\begin{align*}
  \begin{split}
&\left(\frac{\delta\Delta{U}}{k_{\rm B} T} \right)^{2}= \\
&\left(\frac{\partial \Delta{U}}{\partial \tau_{1}}\right)^2 (\delta\tau_{1})^2 + \left(\frac{\partial \Delta{U}}{\partial \kappa_{1}}\right)^2 (\delta\kappa_{1})^2 + \left(\frac{\partial \Delta{U}}{\partial \kappa_{2}}\right)^2 (\delta\kappa_{2})^2 \\
&=(0.2)^2\left(1+\frac{\tau_{1}^2}{\tau_{2}^2}+2\frac{\tau_{1}}{\tau_{2}}\right) + \left(\frac{0.04}{2}\right)^2 + \left(\frac{0.04}{2}\right)^2 \\
&\simeq 0.16.
\end{split}
\end{align*}
Because the propagation of uncertainties is logarithmic for $\delta\Delta{U}$, resolution of energy differences between the two well is lower than half a $k_{\rm B} T$ despite the uncertainty on average lifetimes.

Taking $M = \overline{M}$ (given the small $\delta{M}$ value) allows us computing the sensitivity of $\Delta{U}$ as a function of $\ell$. The computation yields $\partial\Delta{U} / \partial\ell = 0.07\; {k_{\rm B} T \cdot {\rm nm}^{-1}}$. Combining this sensitivity with the uncertainty $\delta\Delta{U}$ of $0.4 \ k_{\rm B} T$ gives an uncertainty on the waist-mirror position $\delta{\ell}$ of $6$ nm only. This rather high spatial resolution is an interesting by-product of our approach.

The reversible heat measured with our method on our experimental configuration through the cross-over path (see main text) is computed from trap stiffnesses which depend on the waist-mirror distance $\ell$. The measured heat uncertainty $\delta{Q_{\rm rev}}$ produced along the path is thus estimated from the determination of $\delta{\ell}$. The trap stiffness of the stable positions (around $z=0$) is computed for an incremental displacement of $\ell$ of $\pm \delta{\ell}$. A worse-case scenario is then followed, taking the highest differences in trap stiffnesses between $\kappa(\ell)$ and $\kappa(\ell+\delta{\ell})$. The heat uncertainty can then be computed as $\delta{Q_{\rm rev}} = \left|k_{\rm B}T\ln \left[\sqrt{ {\kappa(\ell)}/{\kappa(\ell+\delta{\ell})}}\right]\right| \simeq 2.2 \times 10^{-23} \ \mathrm{J}$ with $\kappa(\ell) = 6.47 \times 10^{-6} \ \mathrm{N}\cdot {\rm m}^{-1}$ and $\kappa(\ell+\delta{\ell}) = 6.40 \times 10^{-6} \ \mathrm{N}\cdot {\rm m}^{-1}$. Similarly, the worse-case uncertainty for the  potential energy $\delta{\Delta
U_M = U_M(0,\ell+\delta{\ell}) - U_M(0,\ell)}$ is determined and is about $6 \times 10^{-23} \ \mathrm{J}$.

\end{document}